\documentclass[twocolumn,english]{article}
\usepackage{lmodern}

\usepackage[T1]{fontenc}
\usepackage[latin9]{inputenc}
\usepackage{babel}
\usepackage{array}
\usepackage{verbatim}
\usepackage{rotating}
\usepackage{varioref}
\usepackage{multirow}
\usepackage{amsthm}
\usepackage{amsmath}
\usepackage[unicode=true,
 bookmarks=false,
 breaklinks=true,pdfborder={0 0 0},backref=false,colorlinks=false]
 {hyperref}
\hypersetup{pdftitle={Why Nominal-Typing Matters in Mainstream OOP},
 pdfauthor={Moez A. AbdelGawad},
 pdfsubject={Nominal-Typing in OOP},
 pdfkeywords={Nominal-Typing, OOP, NOOP, Structural-Typing, Java, C\#, Scala}}

\makeatletter

\newcommand{\noun}[1]{\textsc{#1}}
\providecommand{\tabularnewline}{\\}

\numberwithin{equation}{section}
\numberwithin{figure}{section}

\newcommand{\code}[1]{\texttt{#1}}

\makeatother

\usepackage{listings}

\begin{document}

\title{Why Nominal-Typing Matters\\
in Object-Oriented Programming}

\author{Moez A. AbdelGawad\smallskip{}
\\
College of Mathematics and Econometrics, Hunan University\\
Changsha 410082, Hunan, P.R. China\smallskip{}
\\
Informatics Research Institute, SRTA-City\\
New Borg ElArab, Alexandria, Egypt\smallskip{}
\\
\textsf{moez@cs.rice.edu}}
\maketitle
\begin{abstract}
The statements `inheritance is not subtyping' and `mainstream OO
languages unnecessarily place restrictions over inheritance' have
rippled as mantras through the PL research community for years. Many
mainstream OO developers and OO language designers, however, do not
accept these statements. In \emph{nominally-typed} OO languages that
these developers and language designers are dearly familiar with,
inheritance simply is subtyping; and they believe OO type inheritance
is an inherently nominal notion, not a structural one.

Nominally-typed OO languages, such as Java, C\#, C++, and Scala, are
among the most used programming languages today. However, the value
of nominal typing to mainstream OO developers, as a means for designing
robust OO software, seems to be in wait for full appreciation among
PL researchers---thereby perpetuating an unnecessary schism between
many OO developers and language designers and many OO PL researchers,
with each side discounting, if not even disregarding, the views of
the other.

In this essay we strengthen and complement earlier efforts to demonstrate
the semantic value of nominal typing by presenting a technical comparison
between nominal OO type systems and structural OO type systems. Recently,
a domain-theoretic model of nominally-typed OOP was compared to well-known
models of structurally-typed OOP. Combined, these comparisons provide
both a clear technical account and a deep mathematical account of
the relation between nominal and structural OO type systems that have
not been presented before, and they help demonstrate the key value
of nominal typing and nominal subtyping to OO developers and language
designers.

We believe a clearer understanding of the key semantic advantage of
pure nominal OO typing over pure structural OO typing can help remedy
the existing schism. We believe future foundational OO PL research,
to further its relevance to mainstream OOP, should be based less
on structural models of OOP and more on nominal ones instead.
\end{abstract}

\section{Introduction}

\global\long\def\NOOP{\mathbf{NOOP}}
In 1990, Cook et al. shocked the programming languages (PL) research
community by declaring that, in object-oriented programming, `inheritance
is not subtyping,' meaning there is no one-to-one correspondence
between type inheritance and subtyping in OO programming languages,
further adding that mainstream OO languages unnecessarily `place
restrictions over inheritance.' Over the years, these statements
rippled, as mantras, through the PL research community.

To this day, however, many mainstream OO developers and OO language
designers cannot digest or accept not identifying type inheritance
with subtyping. Simply, the statement of Cook et al. is not true in
nominally-typed OO languages that these developers and language designers
are dearly familiar with. Further, they see that the so-called ``restriction''
of type inheritance in nominally-typed OO languages is strongly justified,
or, more so, that a structural view of inheritance is in fact an unjustified
redefinition of type inheritance, which they view as an inherently
nominal notion.

Nominally-typed OO languages are among the top most-used programming
languages today. Examples of nominally-typed OO languages include
many industrial-strength mainstream OO programming languages such
as Java~\cite{JLS14}, \noun{C\#~\cite{CSharp2007},} \noun{C++}~\cite{CPP2011},
and Scala~\cite{Odersky09}. Nominally-typed OO languages have remained
among the top most-used programming languages for over a decade~\cite{Aldrich:2013:PIW:2509578.2514738,TIOBE}.
And, even by the most conservative measures, these languages are expected
to remain among the top most-used OO programming languages in the
near future, if not the far one, too.

In spite of this, the value of nominal typing and nominal subtyping
to mainstream OO developers, as a means for designing robust OO software
that can be readily understood and maintained, as well as the value
of properties of OO type systems that depend on nominality to them
(such as the identification of type inheritance with subtyping), seem
to be not yet fully appreciated among OO PL researchers. This has
led to a continuing tension and schism between two large and significant
communities: many mainstream OO developers and OO language designers,
on one side, and many OO PL researchers, on the other side, with each
of both sides discounting, and even disregarding, the views and opinions
of the other.

In nominally-typed (\emph{a.k.a.}, nominatively-typed) OO languages,
objects and their types are nominal, meaning that objects and their
types carry\emph{ class names information} as part of the meaning
of objects and of the meaning of their types. Class names---and interface
names and trait names in OO languages that support these notions---are
used as type names in nominally-typed OO languages. Class---and interface
and trait---behavioral contracts, typically written informally in
code documentation comments, are specifications of the behavioral
design intentions of OO software developers. In nominally-typed OOP,
a reference to a class (or interface or trait) name is invariably
considered a reference to the associated contract too. Given this
association of type names to corresponding behavioral contracts, \emph{nominal}
\emph{typing allows associating types of objects} \emph{with (formal
or informal) behavioral contracts}.

By using type names in their code, OO developers using nominally-typed
OO languages have a simple way to refer to the corresponding contracts---referring
to them as richer specifications of object state and behavior that
\emph{can} be checked statically and that can be used during runtime.
This readily access to richer object specifications---which cannot
be expressed in a natural way using non-nominal (\emph{a.k.a.}, structural)
record types that, by definition, include no class names information---makes
nominally typed OO languages closer to being semantically typed languages
than structurally typed OO languages are.

The first mathematical models of OOP to gain widespread recognition
among programming languages (PL) researchers were developed while
OOP was making its first steps into mainstream computer programming.
(See Section~\ref{sec:Related-Work}.) These early models were \emph{structural}
models of OOP. As the developers of these models themselves (\emph{e.g.},
Cardelli) explained, this was due to influence from functional programming
research extant at that time. These models of OOP, thus, reflected
a view of OOP that does \emph{not }include class names information.

Being structural, objects were viewed in these models simply as being
records (of functions). Object types, in accordance, were viewed as
record types, where the type of an object only specifies the structure
of the object, meaning that object types carry information only on
the names of the members of objects (\emph{i.e.}, fields and methods)
and, inductively, on the (structural) types of those members. Examples
of structurally-typed OO languages include lesser-known languages
such as O'Caml~\cite{OCamlWebsite}, Modula-3~\cite{Cardelli89modula},
Moby~\cite{Fisher1999}, PolyTOIL~\cite{Bruce2003}, and Strongtalk~\cite{Bracha1993}.\footnote{A discussion of statically-typed versus dynamically-typed OO languages
(including the non-well-defined so-called ``duck-typing''), and
the merits and demerits of each, is beyond the scope of this essay.
The interested reader should check~\cite{Meijer2004}. In this essay,
we focus on nominal and structural \emph{statically}-typed OO languages.} In pure structurally-typed OO languages, class names information
(also called nominal information) is not used as part of the identity
of objects and their types, neither during static type checking nor
at runtime. Accordingly, nominal information is missing in structural
mathematical models of OOP.

The main practical advantage of structural typing over nominal typing
in OO languages seems to be their ``flexibility,'' \emph{i.e.},
the ability in a structurally-typed OO language to have supertypes
get defined ``after the fact'' (\emph{i.e.}, after their subtypes
are already defined). In light of mainstream OO developers of statically-typed
OO languages not adopting structural typing, the ``inflexibility''
of nominally-typed OO languages seems not to be enough justification
for wider use of structural typing, particularly in light of the advantages
of nominal typing we discuss in this essay.

We attempt thus in this essay to further close the gap that exists
between programming language researchers who maintain a structurally-typed
view of OOP (and who believe in conclusions based on this view, such
as inheritance and subtyping not being in one-to-one correspondence)
and mainstream OO software developers and OO language designers who
maintain a nominally-typed view of OO software (and who, accordingly,
reject conclusions based on the structural view) by giving a precise
technical account of the relation between nominal and structural
OO type systems%
\begin{comment}
, since, even though we do not form a final, conclusive judgment on
whether one of the two approaches to OO typing is superior, we believe
it is critical to have a full and deep understanding of the fundamental
technical differences between the two approaches
\end{comment}
. The essay complements the recent mathematical comparison of a nominal
domain-theoretic model of OOP to structural domain-theoretic models
of OOP.

This essay is structured as follows. First, in Section~\ref{sec:Related-Work}
we give some details on the history of modeling OOP, particularly
details relevant to realizing differences between nominal typing and
structural typing and to the development of nominal and structural
models of OOP.

Given that structural typing, more or less, is understood well among
PL researchers, in Section~\ref{sec:Nominally-Typed-OOP-vs.} we
directly demonstrate the value of behavioral contracts and nominal
typing in mainstream nominally-typed OO languages using a comparison,
followed by a discussion of the comparison. In Section~\ref{sub:Contracts-and-LSP}
we first discuss, in some detail, the value of contracts and the value
of identifying inheritance with subtyping in mainstream OO software
design. Then, using code examples, in Section~\ref{sub:Spurious-Subsumption}
and Section~\ref{sub:Inheritance-Is-Not} we compare nominally-typed
OO type systems and structurally-typed ones to more vividly illustrate
the main technical differences between them. We then conclude in Section~\ref{sub:Type-Names,Circularity,Bin-Methods}
by discussing the nominal and structural views of type names and of
recursive-types, and the importance of recursive types in mainstream
OOP. %
\begin{comment}
Based on this discussion, 
\end{comment}

We conclude the essay by summarizing our findings, and making some
final remarks, in Section~\ref{sec:Conclusion}%
\begin{comment}
, then, in Section~\ref{sec:Future-Work}, suggesting some future
work that can be built on top of ideas presented in this essay
\end{comment}
.

\section{\label{sec:Related-Work}Related Work}

Even though object-oriented programming emerged in the 1960s, and
got mature and well-established in mainstream software development
in the 1980s, the differences between nominally-typed and structurally-typed
OO programming languages started getting discussed by PL researchers
only in the 1990s~\cite{Magnusson91,Porter92,Thorup99}. In spite
of an early hint by Cardelli (see below), the value of investigating
nominal typing and nominal subtyping and their value to OO developers
was not appreciated much however---that is, until about a decade later,
around the year 2000.

In the eighties, while mainstream OOP was in its early days, Cardelli
built the first denotational model of OOP~\cite{Cardelli84,Cardelli88}.
Cardelli's work was pioneering, and naturally, given the research
on modeling functional programming extant at that time (which Cardelli
heavily referred to and relied on), the model Cardelli constructed
was a structural denotational model of OOP.\footnote{Quite significantly, Cardelli in fact also hinted at looking for investigating
nominal typing~\cite[p.2]{Cardelli87}. Sadly, Cardelli's hint went
largely ignored for years, and structural typing was rather \emph{assumed}
superior to nominal typing instead, particularly after the publication
of Cook et al.'s and Bruce et al.'s work as we later discuss.}

In the late eighties/early nineties, Cook and his colleagues worked
to improve on Cardelli's model. Unlike Cardelli, Cook et al. emphasized
in their work the importance of self-references in OOP, at the value
level and at the type level. Their research led them to break the
identification of type/interface inheritance with subtyping~\cite{CookDenotational89,Cook1989,CookInheritance90}.\footnote{A discussion of Cardelli's model and of Cook's model and a comparison
of $\NOOP$, a nominal domain-theoretic model of OOP~\cite{NOOP,NOOPsumm},
versus the structural models of Cardelli and Cook is presented in~\cite{AbdelGawad2016,AbdelGawad2015a}.}

In 1994, Bruce and others presented a discussion of the problem of
binary methods in OOP~\cite{BruceBinary94} (a ``binary method''
is a method that takes a parameter or more of the same type as the
class the method is declared in\footnote{See our later discussion of `false/spurious binary methods'---methods
that are mistakenly identified by structural OO type systems as being
binary methods but their semantics are not those of true binary methods.}). Later, Bruce and Simons promoted the structural view of OOP and
conclusions based on it in a number of publications (\emph{e.g.},
\cite{BruceFoundations02} and~\cite{SimonsTheory02}), in spite
of the disagreement between these conclusions and the fundamental
intuitions of a significant portion of mainstream OO developers and
language designers~\cite{InhSubtyNWPT13}.

Under the pressure of this disagreement, some PL researchers then
started in the late nineties/early 2000s stressing the significance
of the differences between nominally-typed OOP and structurally-typed
OOP, and they started acknowledging the practical value of nominal
typing and nominal subtyping~\cite{TAPL}. Accordingly some attempts
were made to develop OO languages with complex type systems that are
both nominally- and structurally-typed~\cite{Findler04,Ostermann08,Gil08,Malayeri08,Malayeri2009,Odersky09,GoWebsite}.
However, in the eyes of mainstream OO developers, these ``hybrid''
languages have more complex type systems than those of languages that
are either simply purely nominally-typed or purely structurally-typed.
This more complexity typically results in lesser productivity for
developers who attempt to use both of the typing approaches in their
software (see also discussion at the end of Section~\ref{sub:Type-Names,Circularity,Bin-Methods}).

As to operational models of OOP, Abadi and Cardelli were the first
to present such a model~\cite{SemObjTypes94,TheoryOfObjects95}.
Again, their model had a structural view of OOP. However, operational
models of nominally-typed OOP got later developed. In their seminal
work, Igarashi, Pierce, and Wadler presented Featherweight Java (FJ)~\cite{FJ/FGJ}
as an operational model of a nominally-typed OO language. Even though
FJ is not the first operational model of nominally-typed OOP (see~\cite{drossopoulou99},~\cite{nipkow98}
and~\cite{flatt98,flatt99}, for example), FJ is the most widely
known operational model of (a tiny core subset of) a nominally-typed
mainstream OO language, namely Java. The development of FJ and other
operational models of nominally-typed OOP motivated the construction
of $\NOOP$ as the first domain-theoretic%
\begin{comment}
 (\emph{a.k.a.}, denotational)
\end{comment}
{} model of nominally-typed OOP~\cite{NOOP,NOOPsumm}.\footnote{Featherweight Java~\cite{FJ/FGJ} offers a very clear operational
semantics for a tiny nominally-typed OO language. It is worth mentioning
that domain-theoretic models of nominally-typed OO languages, as more
foundational models that have fewer assumptions than operational
ones, provide a denotational\emph{ }justification for the inclusion
of nominal information in FJ. The inclusion of nominal information
in $\NOOP$ was crucial for proving\emph{ }the identification of inheritance
with subtyping in nominally-typed OOP~\cite{NOOP,NOOPsumm}. In\noun{
FJ~\cite{FJ/FGJ}}, the identification of inheritance with subtyping
was taken as an\emph{ assumption} rather than being proven as a consequence
of nominality. Also, domain-theoretic models such as $\NOOP$ allows
discussing issues of OOP such as type names, `self-types' and binary
methods on a more foundational level than provided by operational
models of OOP such as FJ. The more abstract description of denotational
models results in a conceptually clearer understanding of the programming
notions described as well as the relations between them.

(It is also worth mentioning that $\NOOP$ was developed, partially,
in response to the technical challenge Pierce presented in his LICS'03
lecture~\cite{Pierce03} in which Pierce looked for the precise relation
between structural and nominal OO type systems, notably\emph{ after}
the development of FJ was concluded, implying that the question about
the relation remained an open question after the development of FJ.
As to their purpose, it is customary that denotational models and
operational ones play complementary roles, where denotational models
are usually of more interest to programming language designers, while
operational ones are usually of more interest to programming language
implementers.)}

Given the different basis for deriving data structuring in functional
programming (based on standard branches of mathematics) and in object-oriented
programming (based on biology and taxonomy\footnote{Which is a fact that seemingly is nowadays forgotten by some PL researchers
but that Cardelli explicitly mentions in~\cite{Cardelli84,Cardelli88}.}), some PL researchers have also expressed dissatisfaction with assuming
that the views of programming based on researching functional programming
(including a view that assumes structural typing) may apply, without
qualifications, to object-oriented programming. In addition to Pierce
and other earlier and later researchers pointing out the importance
of distinguishing between nominal typing and structural typing, MacQueen~\cite{MacQueenMLOO02},
for example, also noted many mismatches between Standard ML (a popular
functional programming language~\cite{Milner97}) and class-based
OO languages such as Java and C++. Later, Cook~\cite{cook-revisited}
also pointed out differences between objects of OOP and abstract data
types (ADTs) that are common in functional programming.

These research results run in a similar vein as ours since they somewhat
also point to some mismatches between the theory and practice of programming
languages---theory being more mathematics-based, functional, and structurally-typed,
and practice being more biology/taxonomy-based, object-oriented, and
nominally-typed.

\section{\label{sec:Nominally-Typed-OOP-vs.}Nominally-Typed OOP versus Structurally-Typed
OOP}

In this section we first informally and less-technically, discuss
the importance of contracts, and of nominal typing and nominal subtyping
to mainstream OO developers in Section~\ref{sub:Contracts-and-LSP}---briefly
discussing DbC (Design by Contract) in the process, then we discuss
how LSP (Liskov's Substitution Principle) expresses the importance
of preserving contracts upon inheritance.

In Section~\ref{sub:Spurious-Subsumption} and Section~\ref{sub:Inheritance-Is-Not},
we then present code examples that illustrate how nominally-typed
OOP and structurally-typed OOP compare to each other from a technical
point of view, and illustrate how structural typing and structural
subtyping sometimes force the breaking of contracts. In the comparison
we discuss two key problems with structural OO type systems, namely,
`spurious subsumption,' and its converse, `missing subsumption.'\footnote{A technical overview of main OO typing notions that explains some
of the technical jargon used in this essay is presented in~\cite{OOPOverview13}.} In Section~\ref{sub:Spurious-Binary-Methods}, we uncover a third
so-far-unrecognized problem in pure structurally-typed OO languages
that we call the problem of `spurious binary methods.'

We then conclude our demonstration of the value of nominal typing
by discussing, in some depth, the nominal and structural views of
type names and of recursive-types, and the importance of recursive
types in mainstream OOP in Section~\ref{sub:Type-Names,Circularity,Bin-Methods}.

The discussions and comparisons in this section demonstrate that nominal
typing in nominally-typed OO programming languages causes typing and
subtyping in these languages to be closer to semantic typing and subtyping,
respectively, because of the association of nominal information with
class contracts. This closeness to semantic typing, the simplicity
of the resulting software design mental model, and the importance
of recursive types to mainstream OO developers and OO language designers
help explain the practical value of nominal typing to mainstream OO
developers and OO language designers.

\subsection{\label{sub:Contracts-and-LSP}Contracts, Nominality and The Liskov
Substitution Principle (LSP)}

Contracts are widely-used notions in mainstream OO software development.
A contract in an OO program is similar to a contract in the real world:
It specifies what an object expects of client objects and what client
objects can expect of it. Members of an object---\emph{i.e.}, its
fields and methods---and properties of these members form the object's
interface with the outside world; the buttons on the front of a television
set, for example, are the interface between us and the electrical
wiring on the other side of the TV's plastic casing. One presses
the ``power'' button and he or she are promised this will turn the
television on and off. In its most common form, an interface is a
group of related methods together with a contract giving promises
on the behavior of these methods. Similarly, a class contract is an
agreement that instances of the class will expose (present as their
public interface, or API) certain methods, certain properties, and
certain behaviors.

\emph{Contract Examples}~~Examples of contracts in OO software are
plenty. Examples familiar to most Java developers, for example, include
the contract of the \code{Comparable} interface promising its clients
a total ordering on its elements and requiring that classes that implement
the interface adhere to this promise, and the contract of class \code{Object}
promising that the \code{equals()} method and the \code{hashCode()}
method are in agreement and requiring subclasses that override one
of the two methods to override the other method accordingly.

Also in Java, class \code{JComponent} contains a default implementation
of all of the methods in the \code{Accessible} interface, but \code{JComponent}
is \emph{not} actually declared to implement the interface, because
the contract associated with \code{Accessible} is \emph{not} satisfied
by the default implementation provided in \code{JComponent}. This
example stresses the association of \emph{inherited} contracts with
superclass names. In mainstream OO software, if a class extends another
class or implements an interface it is declaring that it \emph{inherits
the contract} associated with the superclass or superinterface and
will maintain it. Likewise, if a class does not maintain the contract
associated with another class or interface it does not declare itself
as extending or implementing the class or interface. We discuss this
point further in Section~\ref{sub:Inheritance-and-Contract}.

Other examples of contracts may also include a class that implements
tree layout algorithms. The contract of such a class may require the
input graph to be a tree, and may promise as result, if the input
is a tree, to produce a layout that has no overlapping nodes, edges
or labels.

In general, contracts, whether written formally or informally, usually
contain the following pieces of information: side effects, preconditions,
postconditions, invariants, and, sometimes, even performance guarantees.
In Java, class contracts, as a set of requirements and promises, are
usually stated in Javadoc comments. Requirements of a contract are
simply any conditions on the use of the class, for example: conditions
on argument values, conditions on the order of execution of methods,
conditions on execution in a multi-threaded or parallel environment. 

Two further, rather artificial, examples for contracts, that we will
expound on below to show the differences between nominal typing and
structural typing, are the promise that an animal can play with any
another animal, and the promise that a (mathematical) set contains
no repeated elements. In particular, we use these two examples to
show how structural subtyping can lead to \emph{breaking} contracts
associated with classes/interfaces/traits.

From the presented examples, it is easy to see that a contract is
made of two parts: requirements upon the caller (``the client'')
made by the class (``the provider'') and promises made by the class
to the caller. If the caller fulfills the requirements, then the class
promises to deliver some well-defined service. Requirements may be
enforced by throwing checked or unchecked exceptions when the stated
conditions are violated. Promises can be enforced by assertions at
the end of a method.

Further, according to proponents of `Design By Contract' (DbC),
classes of a software system communicate with one another on the basis
of precisely defined benefits and obligations~\cite{Meyer92,Meyer95}.
If preconditions are not obeyed by the client of the class method,
the service provider will deny its service. If any postcondition\textbf{
}or\textbf{ }invariant is violated, it uncovers a problem on the service
provider side~\cite{Kramer99}. As such, the benefits and obligations
of clients and providers, along with their relative chronological
order, can be summarized as in Table~\ref{tab:Design-By-Contract}.

\begin{table}
\noindent \begin{centering}
\begin{tabular}{|c|l|l|}
\cline{2-3} 
\multicolumn{1}{c|}{} & Benefit & Obligation\tabularnewline
\hline 
\multirow{4}{*}{\begin{turn}{90}
Client
\end{turn}} & (4) Output guaranteed to & \multirow{2}{*}{(1) Satisfy }\tabularnewline
 & comply to postconditions & \tabularnewline
 & (no need to check output) & preconditions\tabularnewline
 &  & \tabularnewline
\hline 
\multirow{4}{*}{\begin{turn}{90}
Provider
\end{turn}} & (2) Input guaranteed to & \multirow{2}{*}{(3) Satisfy}\tabularnewline
 & comply to preconditions & \tabularnewline
 & (no need to check input) & postconditions\tabularnewline
 &   & \tabularnewline
\hline 
\end{tabular}
\par\end{centering}

\protect\caption{\label{tab:Design-By-Contract}Design By Contract (DbC): Benefits
and Obligations {[}Source:~\cite{Kramer99}{]}}
\end{table}

\emph{Nominally-Typed OOP}~~Moving on from DbC to the design of
mainstream OO type systems, OO languages should ideally include behavioral
contracts in object types. Motivated by DbC, contracts are used by
mainstream OO developers for constructing robust, reliable, reusable
and maintainable OO software, since contracts promise specified properties
of objects. For example, in his widely-known book titled `Effective
Java'~\cite{bloch01,bloch08}, Joshua Bloch reflects on the use
of contracts in mainstream OOP and asserts the value of contracts
to OO software design by writing that,
\begin{quotation}
\emph{`No class is an island. Instances of one class are frequently
passed to another. Many classes depend on the objects passed to them
obeying the contracts associated with their superclasses ... once
you\textquoteright ve violated the contract, you simply don\textquoteright t
know how other objects will behave when confronted with your object.'}
\end{quotation}

In practice however, the inclusion of behavioral contracts in object
types is too much to ask of a type checker (because of the general
problem of not being able to statically check contracts, since behavioral
contracts are remarkably expressive). The solution OO language designers
choose is to go with an approximation. The association of class names
with contracts, and OO type systems respecting nominal information
in typing and subtyping decisions, allows a nominally-typed OO type
system to be a \emph{tractable approximation} of DbC; hence, OO language
designers of many mainstream OO languages use nominal typing. Nominally-typed
OO languages typically do not require the enforcement of requirements
and promises of contracts; requirements and promises are rather assumed
to hold, thereby encouraging but not requiring developers to enforce
the contracts. To accurately reflect how contracts are used in nominally-typed
OOP, Table~\ref{tab:Design-By-Contract} can as such be modified
into Table~\ref{tab:Contracts-in-Nominally-Typed}.

\noindent \begin{center}
\begin{table}
\noindent \begin{centering}
\begin{tabular}{|c|l|l|}
\cline{2-3} 
\multicolumn{1}{c|}{} & Benefit & Obligation\tabularnewline
\hline 
\multirow{4}{*}{\begin{turn}{90}
Client
\end{turn}} & (4) Output assumed to & \multirow{2}{*}{(1) Satisfy contr-}\tabularnewline
 & comply to contract & \tabularnewline
 & promises (no need to & act requirements\tabularnewline
 & check output) & \tabularnewline
\hline 
\multirow{4}{*}{\begin{turn}{90}
Provider
\end{turn}} & (2) Input assumed to & \multirow{2}{*}{(3) Satisfy contr-}\tabularnewline
 & comply to contract & \tabularnewline
 & requirements (no need & act promises\tabularnewline
 & to check input) & \tabularnewline
\hline 
\end{tabular}
\par\end{centering}

\protect\caption{\label{tab:Contracts-in-Nominally-Typed}Contracts in Nominally-Typed
OOP: Benefits and Obligations}
\end{table}

\par\end{center}

\subsubsection{\label{sub:Inheritance-and-Contract}Inheritance, Subsumption and
Contract Preservation}

As to the inheritance of contracts in mainstream OOP, implementing
an interface, for example, allows a class in an OO program to become
more formal about the behavior it promises to provide. Interfaces
form a contract between a class and the outside world. In a statically-typed
language the tractable component of the contract is enforced at build
time by the compiler. In Java, for example, if a class claims to implement
an interface, all methods defined by that interface must appear in
its source code before the class will successfully compile, and during
run time it is assumed the promises given by the interface are maintained
by the class. The same happens if a class claims to extend another
class, or if an interface claims to extend another interface. This
inheritance of requirements and promises is sometimes referred to
as \emph{interface inheritance}, \emph{contract inheritance}, or \emph{type
inheritance}.\footnote{According to~\cite{Tempero:2013:PIJ:2524984.2525016}, `Programmers
employ inheritance for a number of different purposes: to provide
subtyping, to reuse code, to allow subclasses to customise superclasses'
behaviour, or just to categorise objects'. Inheritance as only being
`a method by which classes share implementations' (\emph{i.e.},
it being a `code sharing/code reuse' technique), is a very limited
notion of inheritance that, unfortunately, is still entertained by
some OO PL researchers. Code sharing is only a part of the fuller
picture of OO inheritance, and is only a means towards the higher
goal of classes sharing their contracts (and even their architectures~\cite{Fayad:1997:OAF:262793.262798,Aldrich:2013:PIW:2509578.2514738}),
not just their code. Inheritance as `contract sharing' is the notion
of inheritance that we are generally interested in and discuss in
this essay.}

While using inheritance, it may be necessary to make changes to a
superclass contract. Some changes to a specification/contract will
break the caller, and some will not. For determining if a change will
break a caller, professional OO developers use the memorable phrase
``\emph{require no more, promise no less}'': if the new specification
does not require more from the caller than before, and if it does
not promise to deliver less than before, then the new specification
is compatible with the old, and will not break the caller.

Bloch~\cite{bloch01,bloch08}, hinting at the conventional wisdom
among mainstream OO developers that identifies type inheritance with
subtyping, proceeds to conclude, based on his earlier observations
about contracts, that
\begin{quotation}
\emph{`inheritance {[}of contracts{]} is appropriate only in circumstances
where the subclass really is a subtype of the superclass.'}
\end{quotation}
As such, Bloch concludes that 
\begin{quotation}
\emph{`it is the responsibility of any subclass overriding the methods
of a superclass to obey their general contracts; failure to do so
will prevent other classes that depend on the contracts from functioning
properly in conjunction with the subclass.'}
\end{quotation}
The requirement that subclasses maintain the contracts of their superclasses%
\begin{comment}
\footnote{As hinted to earlier, contracts associated with `interfaces' and
`traits,' which are supported in some OO languages, are also included
in our discussion. In this essay we discuss OO languages with multiple
(type) inheritance, not merely ones with single (class) inheritance.
Unless otherwise noted, it should be kept in mind that, generally-speaking,
we use the term `class' to mean `class, or interface, or trait,'
particularly when discussing contracts, types and (type) inheritance.}
\end{comment}
{} is expressed among professional OO developers by stating that well-designed
OO software \emph{should }obey the Liskov substitution principle (LSP)~\cite{liskov87,liskov94}.
According to Bloch~\cite{bloch01,bloch08},
\begin{quotation}
\emph{`The LSP says that }any important property of a type should
also hold for its subtypes\emph{, so that any method written for the
type should work equally well on its subtypes~\cite{liskov87,liskov94}'.}\footnote{The LSP is the third of the five OO design principles (the L in `SOLID')
mainstream OO developers follow to design robust OO software. In the
jargon of OO developers, OO code ``smells'' (in particular, it has
a `refused bequest') if some class in the code does not obey LSP,
i.e., if a derived class\emph{---}that is, a subclass---in the code
breaks the contract of one of its base classes\emph{-}--that is, of
one of its superclasses. The LSP thus expresses that class contracts
are preserved by inheriting classes.}
\end{quotation}
As such, as demonstrated by Bloch, it is common knowledge among professional
mainstream OO developers that subsumption (as expressed by the LSP)
and the identification of inheritance with subsumption (\emph{i.e.},
subtyping between object types) are an integral part of the mental
model of object-oriented programming. Whenever these principles are
violated, a program becomes more difficult to understand and to maintain.
(For example, after his discussion of contracts Bloch then gives examples
demonstrating problems, in the Java libraries which Bloch himself
coauthored, that resulted from violating these principles~\cite{bloch01,bloch08}.)

Based on the discussion of contracts and inheritance, two clear OO
design principles among professional mainstream OO developers are:
\begin{enumerate}
\item Whenever the contract of a class/interface/trait is obeyed by an inheriting
class (\emph{i.e.}, whenever we have type inheritance) we should have
subsumption between corresponding class types (\emph{i.e.}, we should
have subtyping), and, conversely,
\item Whenever we have subsumption between two class types (\emph{i.e.},
whenever we have subtyping) the contracts (of the superclass type)
should be obeyed by the subclass type (\emph{i.e.}, we should have
type inheritance).
\end{enumerate}
Given the importance of the LSP (as expressing the preservation of
contracts upon inheritance) to mainstream OO developers, and given
that contracts are typically specifications of object behavior, it
is easy to conclude that basing typing on contracts so as to make
typing and subtyping closer to behavioral typing and behavioral subtyping
(sometimes also called `semantic typing' and `semantic subtyping')
is a \emph{desirable} property of an OO language.

To further illustrate the importance of identifying inheritance with
subtyping to OO software developers, in the following two sections
we present code examples that point out two problems with structural
subtyping that do \emph{not} exist in nominally-typed OOP. 

The first problem is what is sometimes called the problem of `spurious
subsumption' (\cite[p.253]{TAPL}), and the second is the problem
of inheritance not implying subtyping (which we call `missing subsumption'),
\emph{i.e.}, that inheritance between two classes does not imply subsumption
between the two corresponding class types (the converse of the spurious
subsumption problem.)\footnote{The code examples can be skipped by a reader familiar with these two
problems.}

In the spurious subsumption problem, we have two classes whose instances
do \emph{not} maintain the same contract but are considered subtypes
according to structural subtyping rules, demonstrating an example
of structural subtyping breaking LSP. In the missing subsumption problem,
we have two classes whose instances \emph{do} maintain the same behavioral
contract but that are not considered subtypes in a structural type
system, due to structural type systems rebinding self-types upon inheritance,
demonstrating an example of structural subtyping thus breaking the
identification between inheritance and subtyping.

Further, due to pure structural OO type systems always requiring the
rebinding of self-types upon inheritance, we also point out a so-far-unknown
problem with pure structural subtyping that we call the problem of
`spurious binary methods.'

\subsection{\label{sub:Spurious-Subsumption}Spurious Subsumption}

In structurally-typed OOP, subtyping does not imply inheritance, that
is, we may have subtyping between types corresponding to two classes
(instances of one can be used as that of the other) but not have an
inheritance relation between these two classes. To illustrate, let
us assume the following definitions for class \code{Set} and class
\code{MultiSet} (where the contract of class \code{Set} disallows
repetition of elements of a \code{Set}, in agreement with the mathematical
definition of sets, whereas the contract of class \code{MultiSet}
allows repetition of elements of a \code{MultiSet}, in agreement
with the mathematical definition of multisets, sometimes also called
bags),\textbf{}
\begin{lstlisting}[basicstyle={\small\ttfamily},language=Java]
  class Set {
    Boolean equals(Object s) { ... }
    Void insert(Object o) { ... }
    Void remove(Object o) { ... }
    Boolean isMember(Object o) { ... }
  }

  class MultiSet {
    Boolean equals(Object ms) { ... }
    Void insert(Object o) { ... }
    Void remove(Object o) { ... }
    Boolean isMember(Object o) { ... }
  }
\end{lstlisting}

\noindent We note in the code for class \code{\noindent Set} and
class \code{\noindent MultiSet} that the two classes support precisely
the same set of four operations on their instances having the same
signatures for these four operations. The \emph{different} contracts
associated with the two classes, specifying that the semantics and
run-time behavior of their instances should agree with the corresponding
mathematical notions, are reflected only\emph{ }in the different class
names of the two classes but not in the structure of %
\begin{comment}
\noindent (\emph{i.e.}, here, the set of operations supported by) 
\end{comment}
the two classes.

Given that nominally-typed OOP respects class names (and thus the
associated class contracts) while structurally-typed OOP ignores them,
we have the final assignment in
\begin{lstlisting}[basicstyle={\small\ttfamily},language=Java]
  MultiSet m = new MultiSet();
  m.insert(2);
  m.insert(2);
  Set s = m; // Allow assignment?
\end{lstlisting}
correctly disallowed in nominally-typed OOP, but is wrongly allowed
in structurally-typed OOP. The assignment is not allowed in nominally-typed
OOP because class \code{MultiSet} does not inherit from class \code{Set},
while it is allowed in structurally-typed OOP because of matching
signatures of all operations supported by the two classes. The assignment
should not be allowed because, if allowed (as demonstrated in the
code above), it will allow repetition of a set's elements in the
value bound to variable \code{s} by the assignment (given that it
is an instance of \code{MultiSet}). Variable \code{s}, by its declaration,
is assumed to be a \code{Set} (with no repeated elements), and the
assignment, if allowed, will thus \emph{break} \emph{the contract
}of class \code{Set} associated with variable \code{s}.

The problem of spurious subsumption is similar to the problem of accidentally
mistaking values of a datatype for those of another datatype (think
of using floats for modeling euros and dollars then mistaking euros
for dollars or mistaking floats for either~\cite{TAPL}.)%
\begin{comment}
, which, reportedly, was the problem responsible for the Apollo disaster~{[}??{]}
\end{comment}
{} Similarly, OO software developers think of an object in the context
of its class hierarchy and of the contracts associated with its class
members. A key prescript of nominally-typed OOP is that class contracts
are inherited along with class members and should be maintained by
instances of inheriting subclasses. Spurious subsumption in structurally-typed
OOP allows for unintended breaking of this rule, since a structural
type checker fails to reject a program that uses an object of one
type where a behaviorally different, but structurally compatible,
type is expected.\footnote{For example, as mentioned earlier, interface \code{Comparable} in
Java, consisting of the single abstract method \code{int compareTo(Object o)},
has a public contract asserting that \code{compareTo} defines a \emph{total
ordering} on instances of any class inheriting from \code{Comparable},
and that clients of \code{Comparable} can thus depend on this property.
An arbitrary class with a method \code{int compareTo(Object o)},
however, generally-speaking does not necessarily obey this contract.
In structurally-subtyped OO languages, instances of such a class can
be bound, by spurious subsumption, to variables of type \code{Comparable}
(similar to allowing the binding of the instance first bound to variable
\code{m} to variable \code{s} in the \code{Set/MultiSet} example
above). This is in spite of the fact that when a developer asserts
that a class \code{C} implements \code{Comparable}, he or she is
asserting that \code{compareTo()} defines a total ordering on class
\code{C}, and the author of \code{Comparable} conversely asserting
that if a class \code{C} does \emph{not} implement the \code{Comparable}
interface, instances of \code{C} \emph{cannot} be bound to variables
of type \code{Comparable}, since the \code{compareTo} method of
\code{C} may unintentionally or intentionally not define a total
ordering on instances of \code{C}.}.

\subsection{\label{sub:Inheritance-Is-Not}Inheritance Is Not Subtyping}

Another problem with structural subtyping is the converse of the spurious
subsumption problem. In structurally-typed OO languages, structural
subtyping does not require subtyping between types of classes that
are in the inheritance relation (and thus have inherited contracts),
\emph{i.e.}, inheritance does not imply subtyping. In combination
with subtyping not implying inheritance (\emph{i.e.}, spurious subsumption),
structural subtyping thus totally separates the notions of inheritance
and subtyping, based on its non-nominal view of inheritance which
\emph{ignores} the inheritance of class contracts associated with
class names.

To illustrate inheritance not implying subtyping in structurally-typed
OOP, assume the following definitions for class \code{Animal} and
class \code{Cat}\textbf{}
\begin{lstlisting}[basicstyle={\small\ttfamily},language=Java]
  class Animal {
    Void move(Point to) { ... }
    Void eat(Food some) { ... }
    Void breathe() { ... }
    Void sleep(Time period) { ... }
    ... // more generic animal behavior,
	// and ...
    Animal mate(Animal a) { ... }
  }

  class Cat extends Animal {
    ... // behavior specific to cats
    
    // generic animal behavior inherited
    // from class Animal

    // but, do Cats...
    Cat mate(Cat c) { ... }
    // OR do they...
    Animal mate(Animal a) { ... } // ???
  }
\end{lstlisting}
Structurally-typed and nominally-typed OOP disagree on the signature
of method \code{mate()} in class \code{Cat}. Structurally-typed
OOP assumes that \code{mate()} is a binary method (See~\cite{BruceBinary94},
where binary methods were recognized as problematic and multiple approaches
were suggested for dealing with them. We discuss them more technically
in Section~\ref{sub:Type-Names,Circularity,Bin-Methods}), and requires
the method to have the ``more natural'' signature \code{Cat mate(Cat~c)},
at the expense of making \code{Cat}s (\emph{i.e.}, instances of class
\code{Cat}) not be \code{Animal}s (\emph{i.e.}, instances of class
\code{Animal})\footnote{That is, in structurally-typed OO languages the structural type corresponding
to class \code{Cat} is not a subtype of the structural type corresponding
to class \code{Animal} (because of the contravariance of types of
method arguments), unless some unintuitive structural notion, like
`matching'~\cite{BruceFoundations02} (which expresses `the similarity
of recursive structure' between class \code{Cat} and class \code{Animal},
but which did not gain traction or support in mainstream OOP), is
added to the language to be used as a \emph{pseudo}-replacement for
subtyping. (See also discussions in Section~\ref{sub:Type-Names,Circularity,Bin-Methods}
and in~\cite{AbdelGawad2016,AbdelGawad2015a})}, when they (quite naturally) should be ones\footnote{One may here recall Cardelli's noting, in~\cite{Cardelli84,Cardelli88},
of the `biological origin' of OOP. This biological origin is the
reason OO inheritance is called `inheritance' in the first place.}.

Nominally-typed OOP, on the other hand, does not assume the \code{mate()}
method in class \code{Animal} to be a binary method. It thus keeps
using the same signature for the method upon its inheritance by class
\code{Cat}. In nominally-typed OOP, thus, \code{Cat}s are indeed
\code{Animal}s, meaning that nominally-typed mainstream OOP does
identify inheritance with subtyping. Given how nominally-typed OOP
and structurally-typed OOP differ on whether inheritance implies subtyping,
the assignment \code{Animal a = new Cat()} is correctly allowed in
nominally-typed OOP, but is wrongly disallowed in structurally-typed
OOP.

To summarize, the code examples presented above demonstrate a fundamental
difference between structurally-typed OOP and nominally-typed OOP
from the perspective of OO developers. In structurally-typed OOP,
where only class structure is inherited but not class contracts, \code{MultiSet}s
are \code{Set}s (contrary to their mathematical definition) and \code{Cat}s
are not \code{Animal}s (contrary to their ``biological definition'').
In nominally-typed OOP, where class contracts are inherited (via class
names) in addition to class structure, \code{MultiSet}s are not \code{Set}s
(in agreement with their mathematical definition) and \code{Cat}s
are \code{Animal}s (in agreement with their ``biological definition''.)

In nominally-typed OOP whether subtyping is needed or not is indicated
by the presence or absence of explicit inheritance declarations.
Accordingly, the code examples above make it clear, more generally,
that:

\emph{Structurally-typed OOP sometimes forces subtyping when it is
unneeded, and sometimes bars it when it is needed, while nominally-typed
OOP only forces subtyping when it is explicitly needed, and bars it
when, by omission, it is explicitly unneeded.}

This conclusion demonstrates a fundamental semantic and practical
value of nominal information to OO developers of nominally-typed OO
programming languages.

\subsubsection{\label{sub:Spurious-Binary-Methods}Spurious Binary Methods}

A lesser-recognized problem with structurally-typed OOP, which is
also related to binary methods, is the fact that, in a pure structurally-typed
OO language (\emph{i.e.,} one with no nominal typing features), a
class like \code{Animal} cannot have a method like, say,

\code{Void playWith(Animal~a)}\\
that keeps having the \emph{same} signature in its subclass \code{Cat}\footnote{So as to allow \code{Cat}s to play with \code{Dog}s and \code{Mouse}s
(\emph{i.e.}, mice!), for example.} and any other subclasses of class \code{Animal}. Pure structurally-typed
OOP \emph{always} treats as a binary method any method inside a class
that takes an argument that has the same type as that of the class
the method is declared in. This causes methods like \code{playWith}
in class \code{Animal}, whose semantics is \emph{not} that of true
binary methods, to be mistaken as ones, and thus, in subclass \code{Cat},
for example, the \code{playWith} method will have the restrictive
signature

\code{Void playWith(Cat~a)}\\
(allowing \code{Cat}s to play only with other \code{Cat}s but not
with other \code{Animal}s.)

We call this so-far-unrecognized problem with structural typing as
the problem of `spurious binary methods' (or, `false\emph{ }binary
methods'), since a method is inadvertently considered as being a
binary method when it should not be. Nominally-typed OO languages
do not suffer from this problem, because nominally-typed OO languages
treat any such method %
\begin{comment}
(\emph{i.e.}, any method that takes an argument that has the same
type as that of the class the method is declared in) 
\end{comment}
as a regular (\emph{i.e.}, non-binary) method, and thus the signature
of the method does not change upon inheritance in a nominally-typed
OO language.\footnote{In nominally-typed OOP, `F-bounded Generics' (as used, for example,
to define the generic class \code{Enum} in Java~\cite{JLS05,JLS14})
offers a somewhat better alternative---if also not a fully satisfactory
one---to support true binary methods, while keeping the identification
between inheritance and subtyping. Based on some preliminary research
we made, we expect future research to offer a more satisfactory alternative
for supporting true binary methods in nominally-typed OO languages---hopefully
a fully satisfactory alternative. (Also see related discussion close
to the end of Section~\ref{sub:Type-Names,Circularity,Bin-Methods}.)}

\subsection{\label{sub:Type-Names,Circularity,Bin-Methods}Nominal Typing, Type
Names, Recursive Types and Binary Methods}

Based on the discussion in the previous sections, a fundamental
technical difference between nominally-typed OO languages and structurally-typed
OO languages clearly lies in how the two approaches of typing OO languages
differently view and treat \emph{type names}.%
\begin{comment}
As we have seen, this difference, in turn, is concretely reflected
in how the two approaches differ in their support of self-referential
classes, and, in particular, of binary methods.
\end{comment}

In structurally-typed OO languages, type names are viewed as being
names for type variables that abbreviate type expressions (\emph{i.e.},
are ``shortcuts''). As such type names in structurally-typed OO
languages are useful, and are even necessary for defining recursive
type expressions. As variable names, however, recursive type names
in structurally-typed OO languages (such as the name of a class when
used inside the definition of the class, which gets interpreted as
`self-type') get \emph{rebound} to different types upon inheritance,
and they get rebound to types that, if they were subtypes, could break
the contravariant subtyping rule of method parameter types and thus
break the type safety of structurally-typed OO languages. Structurally-typed
OO languages resolve this situation by breaking the one-to-one correspondence
between inheritance and subtyping as we demonstrated in the earlier
code examples in Section~\ref{sub:Spurious-Subsumption} and Section~\ref{sub:Inheritance-Is-Not}.

In nominally-typed OO languages, however, nominality of types means
type names are viewed as part of the identity and meaning of type
expressions given the association of type names with public class
contracts. This means that class names \emph{cannot} be treated as
variable names. Accordingly, in a nominally-typed OO program type
names have \emph{fixed} meanings that do not change upon inheritance.
Further, the fixed type a type name is bound to in a nominally-typed
OO program does not break the contravariant subtyping of method parameters
when the method and its type get inherited by subclasses, thus not
necessitating breaking the identification between inheritance and
subtyping as we demonstrated by the earlier code examples in Section~\ref{sub:Spurious-Subsumption}
and Section~\ref{sub:Inheritance-Is-Not}%
\begin{comment}
 in Section~\ref{sub:Inheritance-Is-Not}
\end{comment}
.

In class-based OOP, a class can directly refer to itself (using class
names) in the signature of a field, or that of a method parameter
or return value. This kind of reference is called a \emph{self-reference},
a \emph{recursive reference}, or, sometimes, a \emph{circular} \emph{reference}.
Also, mutually-dependent classes, where a class refers to itself indirectly
(\emph{i.e.}, via other classes), are allowed in class-based OOP.
This kind of reference inside a class, indirectly referencing itself,
is called an \emph{indirect self-reference,} a \emph{mutually-recursive
reference}, or, a \emph{mutually-circular reference}.

As Pierce~\cite{TAPL} noted, nominally-typed OO languages allow
readily expressing circular (\emph{i.e.}, mutually-dependent) class
definitions. Since objects in mainstream OOP are characterized as
being self-referential values (are `self-aware,' or `autognostic'
according to Cook~\cite{cook-revisited}), and since self-referential
values can be typed using recursive types~\cite{MPS}, there is a
strong and wide need for circular class definitions in OOP. As such,
direct and indirect circular type references are quite common in mainstream
OOP~\cite{cook-revisited}. The ease by which recursive typing can
be expressed in nominally-typed OO languages is one of the main advantages
of nominally-typed OOP. According to Pierce~\cite[p.253]{TAPL},
\begin{quotation}
\emph{`The fact that recursive types come essentially for free in
nominal systems is a decided benefit {[}of nominally-typed OO languages{]}.'}
\end{quotation}
As a demonstration of the influence of views of self-referential
classes on properties of OO type systems, when nominal and structural
domain-theoretic models of OOP are compared (as done in brief, \emph{e.g.},
in~\cite{NOOP,NOOPsumm}, and in detail, \emph{e.g.}, in~\cite{AbdelGawad2016}),
it is easy to see that self-referential classes are viewed differently
by nominal models of OOP than by structural models and that these
different views of self-referential classes, in particular, make nominal
domain-theoretic models of OOP lead to a simple mathematical proof
of the identification between type inheritance and subtyping---a different
conclusion than the one reached based on structural models. (In particular,
the inclusion of class signature constructs in $\NOOP$ led to the
simplicity of the mathematical proof of this identification. See~\cite{NOOP,NOOPsumm,AbdelGawad2015a}
for more details.)

Aside from theory, the difference between the nominal and the structural
views of type names in OOP demonstrates itself most prominently, in
practice, in the different support and the different treatment provided
by nominally-typed OO languages and by structurally-typed OO languages
to ``binary methods''%
\begin{comment}
 as we demonstrated in Section~\ref{sub:Inheritance-Is-Not}
\end{comment}
. As mentioned in Section~\ref{sub:Inheritance-Is-Not}, a ``binary
method'' is a method that takes a parameter or more of the same type
as the class the method is declared in~\cite{BruceBinary94}. The
problem of binary methods and requiring them to be supported in OO
languages was a main factor behind structural models of OOP leading
to not identifying type inheritance with subtyping%
\begin{comment}
 (\emph{i.e.}, as not being in a one-to-one correspondence)~\cite{CookInheritance90}
\end{comment}
. Structurally-typed OO languages, given their view of type names
as type variable names that can get rebound, require the type of the
argument of a method, when identified as a binary method and upon
inheritance of the method, to be that of the type corresponding to
the \emph{subclass}. 

Nominally-typed OO languages, on the other hand, with their fixed
interpretation of type names, offer a somewhat middle-ground solution
between totally avoiding binary methods and overly embracing them
as pure structurally-typed OO languages do. Nominally-typed OO languages
treat a method taking in an argument of the same class as that in
which the method is declared like any other method, needing no special
treatment. Nominally-typed OOP thus does not quite support binary
methods, but, for good reasons (\emph{i.e.}, so as to not break the
identification of inheritance of contracts with subtyping nor lose
other advantages of nominal typing), offers only a good \emph{approximation}
to binary methods. Given that the meaning of types names in nominally-typed
OO languages does not change upon inheritance, these languages provide
methods whose type, upon inheritance, only approximates that of true
binary methods: The type of the input parameter of a method that approximates
a binary method is guaranteed to be a \emph{supertype} of its type
if it were a true binary method. Given that the type of the parameter
does not change in subclasses, the degree of approximation gets lesser
the deeper in the inheritance hierarchy the method gets inherited.

In light of the `spurious binary methods' problem we uncovered in
structural OO type systems (see Section~\ref{sub:Inheritance-Is-Not}),
we believe providing approximations to binary methods is a smart design
choice by nominal OO type systems, even if it is likely that avoiding
spurious binary methods may have not been consciously intended. It
should also be noted that the problem of spurious binary methods provides
justification for nominally-typed OO languages being cautious about
fully embracing binary methods by treating a method that looks like
a binary method as indeed being one.

Also, as we hinted to in earlier sections, in our opinion structural
typing having arguably better support for binary methods does not
justify using structural OO typing, since structural type systems
have their own problems in their support of binary methods (\emph{i.e.},
the problem of spurious binary methods). We conjecture that F-bounded
generics, or some other notion\footnote{For example, ``\emph{implicit} self-type-variables'', \emph{e.g.},
\code{This/Self}, which are implicit type variables because they
do not get included in class signatures, similar to ``implicit self-variables'',
\emph{e.g.}, \code{this/self}, which are not included in method signatures.} may provide a better solution for binary methods in nominally-typed
(and possibly also in structurally-typed OOP) that does not require
breaking the identification of inheritance with subtyping (and thus
does not require sacrificing the closeness of nominal typing/subtyping
to semantic and behavioral typing/subtyping and other advantages of
nominal typing). In light of the `spurious binary methods' problem,
and requiring no explicit use of generics, in our opinion a better
approach towards supporting binary methods in mainstream OO languages
might be by allowing developers to explicitly mark or flag binary
methods as being such, or, even more precisely, to allow developers
to \emph{mark specific parameters} of methods as being parameters
that need to be treated as those of binary methods.%
\begin{comment}
 More details on the differences between nominal and structural mathematical
models of OOP are presented in~\cite{AbdelGawad2016,AbdelGawad2015a}.
\end{comment}

As to ``hybrid'' OO languages, which add (some) structural typing
features to nominally-typed OO languages\footnote{Due to problems with supporting recursive types mentioned above, we
believe most of these ``hybrid'' languages do \emph{not} support
recursive structural types, for example.}, or vice versa, we conjecture that the useful part of the claimed
``flexibility'' of structural typing may be possible to achieve
in nominally-typed OO languages by supporting a separate notion of
`contract names,' thereby splitting class names from contract names,
then allowing classes to additionally define themselves as satisfying
supercontracts of other already-defined (sub)classes. We have not
explored this suggestion further however, since we believe it may
complicate nominal OO type systems. But we believe that, if flexibility
is an absolute necessity, then splitting class names from contract
names may be a suggestion worthy of investigation. We believe this
suggestion to be a more viable option---simpler to reason about and
more in agreement with the nominal spirit of nominally-typed OO languages---than
using multiple dispatch (\cite{Chambers92,Boyland97,Clifton06}),
which was discussed in~\cite{BruceBinary94} as a possible solution
to the problem of binary methods%
\begin{comment}
 (see discussion in Section~\ref{sub:Inheritance-Is-Not})
\end{comment}
, and also more viable than creating hybrid languages. We believe
that having an OO type system be both nominally and structurally typed,
as in~\cite{Findler04,Ostermann08,Gil08,Malayeri08,Malayeri2009,Odersky09,GoWebsite},
makes the type system very complex (and probably even lends its ``hybrid''
features unusable.)

\section{\label{sec:Conclusion}Concluding Remarks}

In this essay we added to earlier efforts that aimed to demonstrate
the semantic value of nominal typing, particularly the association
of class names with behavioral class contracts, by making a technical
comparison between nominal OO type systems and structural OO type
systems. Recently, a domain-theoretic model of nominally-typed OOP,
namely $\NOOP$, was also compared to models of structurally-typed
OOP%
\begin{comment}
 developed and disseminated by Cardelli, Cook, and others
\end{comment}
. These comparisons provide a clear and deep account for the relation
between nominal and structural OO type systems that, due to earlier
lack of a domain-theoretic model of nominally-typed OOP, has not been
presented before, and they should help further demonstrate, to OO
PL researchers, the value of nominal typing and nominal subtyping
to mainstream OO developers and language designers, and instill in
them a deeper appreciation of it.

In the essay we particularly noted that \emph{nominal typing} prevents
types that structurally look the same from being confused as being
the same type. Since some objects having the same structure does not
necessarily imply these objects have the same behavior, nominal typing
identifies types only if they have the same class names information
(nominal information) and thus only if they assert maintaining the
same contract and not just assert having the same structural interface.
Thus, in nominally-typed OOP objects having the same class type implies
them asserting they maintain the same contracts, and them asserting
they maintain the same contracts implies them having the same type.

Similarly, \emph{nominal subtyping} allows subtyping relations to
be decided based on the refinement of contracts maintained by the
objects, not just based on the refinements of their structure. By
inclusion of contracts in deciding the subtyping relation, nominal
subtyping thus also prevents types that are superficially (\emph{i.e.},
structurally) similar from being confused as being subtypes. Since
the similarity of structure does not necessarily imply the similarity
of behavior, in nominally-typed OOP type inheritance implies refined
contracts, and refined contracts imply subsumption between class types,
and vice versa (\emph{i.e.}, in nominally-typed OOP, subsumption between
class types implies refined contracts, implying type inheritance.)

Putting these facts together, it is clear that in nominally-typed
OOP different class names information implies different contracts/types
and different contracts/types imply different class names information.
This identification of types with contracts, and of subtyping with
inheritance of contracts, makes nominal typing and nominal subtyping
closer to semantic typing and semantic subtyping.

In the essay we thus stressed the practical value of nominal typing
in mainstream OOP by particularly showing
\begin{enumerate}
\item The value of nominal subtyping, at compile time and at runtime, in
\textbf{respecting} \textbf{behavioral} \textbf{contracts} and thus
respecting design intents,
\item The value of the resulting identification between inheritance and
subtyping in providing a simpler conceptual model of OO software and
of OO software components, leading to a \textbf{simpler} \textbf{design}
\textbf{process} of OO software, and
\item The value of making \textbf{recursive} \textbf{types} \textbf{readily}
\textbf{expressible}, this being necessary for the static typing of
``autognostic'' objects.
\end{enumerate}
Our comparison also revealed the problem of `spurious binary methods,'
so far an unrecognized problem in structurally-typed OO languages.

Further, the recent comparison of nominal and structural denotational
models of OOP%
\begin{comment}
 (\emph{e.g.}, as in~\cite{AbdelGawad2016,AbdelGawad2015a})
\end{comment}
{} demonstrates them having different views of \emph{fundamental} notions
of mainstream OOP, namely of objects, of type names, of object/class
types, of subtyping and of the relation between subtyping and type
inheritance. In particular, this comparison demonstrates that an object
in mainstream nominally-typed OOP is a record together with nominal
information, that class types are record types whose elements (\emph{i.e.},
objects/class instances) additionally respect statements of class
contracts, and that type inheritance is correctly identified with
nominal subtyping.%
\begin{comment}
In addition, the comparison of denotational models of OOP shows that
nominally-typed models and structurally-typed models of OOP lead to
different views of fundamental notions of mainstream OOP, namely objects,
type names, class types, subtyping and the relation between subtyping
and inheritance. In particular, the comparison demonstrates that
\begin{enumerate}
\item An object in mainstream nominally-typed OOP should not be mathematically
viewed as merely denoting a record of its members (\emph{i.e.}, its
fields and methods) but rather as a record together with nominal information
that is associated with class contracts that the object maintains
carried along with the record and constraining its members,
\item Class types should not be viewed as being record types but rather
as record types whose elements (\emph{i.e.}, instances) additionally
respect statements of class contracts, and also that
\item In nominally-typed OOP, type inheritance should be viewed as being
correctly identified with nominal subtyping (\emph{i.e.}, in nominally-typed
OOP, inheritance is subtyping.)\end{enumerate}
\end{comment}

Table~\vref{tab:Nominally-OO-Typing} summarizes the main differences
between nominal typing and structural typing we pointed out in this
essay.

\noindent 
\begin{table*}
\noindent \begin{centering}
\begin{tabular}{|c|c|c|}
\cline{2-3} 
\multicolumn{1}{c|}{} & \noun{Nominally-Typed OOP} & \noun{Structurally-Typed OOP}\tabularnewline
\hline 
\multirow{2}{*}{\textsl{Object Interfaces}} & \multirow{2}{*}{Nominal; Include class names} & Structural; Do not\tabularnewline
 &  &  include class names\tabularnewline
\hline 
\textsl{Contracts} & Included (via class names) in & Ignored in the meaning\tabularnewline
\textsl{(as richer specifications} & the meaning of objects, of their &  of objects, of object interfaces,\tabularnewline
\textsl{of object behavior)} &  interfaces and of their types &  and of object types\tabularnewline
\hline 
\textsl{Types of Objects} & Class types & Record types\tabularnewline
\hline 
\textsl{Object Type} & \multirow{2}{*}{Class signatures} & \multirow{2}{*}{Record type expressions}\tabularnewline
\textsl{Expressions} &  & \tabularnewline
\hline 
\multirow{2}{*}{\textsl{Type Reification/Nominality}} & Class signatures, as object types, & Objects do not carry their types\tabularnewline
 & are included in object meanings &  as part of object meanings\tabularnewline
\hline 
\multirow{3}{*}{\textsl{Type Names}} & Class names, being associated & Type names are abbreviations/\tabularnewline
 & with public class contracts, & synonyms for type expressions\tabularnewline
 &  are used as type names &  (with no inherent fixed meaning)\tabularnewline
\hline 
\textsl{Meaning of } & Fixed. Cannot be & Can be rebound\tabularnewline
\textsl{ Type Names} &  rebound upon inheritance &  upon inheritance\tabularnewline
\hline 
\multirow{2}{*}{\textsl{(Type) Inheritance}} & Includes inheritance of & \multirow{2}{*}{Ignores behavioral class contracts}\tabularnewline
 &  class contracts & \tabularnewline
\hline 
\multirow{2}{*}{\textsl{Subtyping}} & Respects contracts; & \multirow{2}{*}{Ignores behavioral class contracts}\tabularnewline
 &  respects (type) inheritance & \tabularnewline
\hline 
\textsl{(Type) Inheritance} & \multirow{2}{*}{One-to-one correspondence} & \multirow{2}{*}{Two relations independent}\tabularnewline
\textsl{ versus Subtyping} &  & \tabularnewline
\hline 
\textsl{OO Software Design} & Simple; Inheritance hierarchy = & Complex; An inheritance hierarchy and \tabularnewline
\textsl{Mental Model} & Subtyping hierarchy &  a separate, independ't subtyping hierarchy\tabularnewline
\hline 
\multirow{2}{*}{\textsl{Binary Methods}} & Not supported. & Fully supported\tabularnewline
 & Approximations provided &  (including \emph{false} ones)\tabularnewline
\hline 
\textsl{Spurious and Missing} & \multirow{2}{*}{Neither can exist} & \multirow{2}{*}{Both can exist}\tabularnewline
\textsl{ Subsumption} &  & \tabularnewline
\hline 
\textsl{Spurious Bin. Methods} & Cannot exist & Can exist\tabularnewline
\hline 
\multirow{2}{*}{\textsl{Recursive Types}} & \multirow{2}{*}{Readily and naturally expressed} & Special constructs needed for\tabularnewline
 &  & explicit expression\tabularnewline
\hline 
\multirow{2}{*}{\textsl{Typing and Subtyping}} & Closer to semantic/behavioral & Further from semantic/behavioral\tabularnewline
 & typing and subtyping & typing and subtyping\tabularnewline
\hline 
\end{tabular}
\par\end{centering}

\noindent \centering{}\protect\caption{\label{tab:Nominally-OO-Typing}OO Nominal-Typing vs. OO Structural-Typing}
\end{table*}

We hope the development of mathematical models of nominally-typed
OOP and the comparisons presented in this essay and elsewhere are
significant steps in providing a full account of the relation between
nominal and structural OO type systems. We further hope this essay
clearly explains the rationale behind our belief that the significant
practical value and the significant semantic value of nominal typing
are the reasons for industrial-strength mainstream OO software developers
correctly choosing to use nominally-typed OO languages. We believe
that having a clear view of the rationale behind many OO developers'
preference of nominally-typed OO languages, and having a more accurate
technical and mathematical view of nominally-typed OO software, present
programming languages researchers with better chances for progressing
mainstream OO languages and for making PL research relevant to more
OO language designers and more mainstream OO software developers.%
\begin{comment}
\label{sec:Future-Work}Future Work
\end{comment}
\begin{comment}

\paragraph{Which is better for OO type systems, nominal typing or structural
typing?}

As hinted to earlier in this essay, even though we are biased towards
nominal typing and against structural typing (this essay, we hope,
presents some justification for our stance), we do not provide a final,
conclusive answer to this important question.

However, it is worthy here that we note some of our thoughts on answering
this question. 
\end{comment}
\begin{comment}
The main practical advantage of structural typing over nominal typing
in OO languages seems to be their ``flexibility'', \emph{i.e.},
the ability in a structurally-typed OO language to have supertypes
get defined ``after the fact'' (\emph{i.e.}, after their subtypes
are already defined)~\cite{Malayeri08,Malayeri2009}.
\end{comment}
\begin{comment}
 In light of mainstream OO developers not adopting structural typing,
we do not see the ``inflexibility'' of nominally-typed OO languages,
however, as enough justification for using structural typing, particularly
in light of the presented advantages of nominal typing to mainstream
OO developers we discussed in this essay. 
\end{comment}
\footnote{Generics, for example, add to the expressiveness of type systems
of nominally-typed OO programming languages~\cite{Bank96,Bracha98,Corky98,JLS05,JLS14,CSharp2007,bloch08,ScalaWebsite,GenericsFAQWebsite,Zhang:2015:LFO:2737924.2738008}.
As hinted to earlier, generics---particularly so-called `F-bounded
generics'---seem to improve the support for binary methods in nominally-typed
OO languages  while maintaining the identification between inheritance
and subtyping. We believe building a domain-theoretic model of generic
nominally-typed OOP (\emph{e.g.}, along the lines of~\cite{AbdelGawad2016a})
may offer better chances for having a deeper understanding of features
of generic mainstream OO languages, such as Java erasure, variance
annotations (including the notorious Java wildcards), polymorphic
methods, generic type inference and so on.}

\begin{comment}
Further, having better appreciation of the value of nominal typing
may lead to a better understanding of features of mainstream OOP that
seem to depend on nominality. As an example, it has been suggested~\cite{AbdelGawad2014b}
that Java wildcards (as a form of variance annotations) might be best
understood as intervals (over subtyping partially-ordered sets, \emph{i.e.},
subtyping posets) and that, after the addition of generics and wildcards
in particular, the structure of the graph of the subtyping relation
in Java may be best understood not merely as a directed-acyclic graph
(DAG) but, more specifically, as a fractal. This observation has been
suggested to possibly offer better chances for improving our understanding
of Java generics, thereby offering chances to improve Java compilers,
\emph{e.g.}, improving the compilers' diagnostic error messages related
to generics and wildcards. As mentioned in~\cite{AbdelGawad2014b},
the nominality of typing and of subtyping in Java plays a prominent
role in the intervals model of Java wildcards and in viewing the graph
of subtyping in Java as a fractal.
\end{comment}

Finally, we believe a clearer understanding and a deeper appreciation
of a key semantic advantage of nominal OO typing over structural OO
typing %
\begin{comment}
\textemdash as provided in this essay, via appreciating the association
of class names to richer object specifications\textemdash{}
\end{comment}
can help remedy the existing schism between OO PL researchers on one
hand and OO developers and OO language designers on the other hand,
offering thereby better chances for progressing mainstream OO languages.
In particular, we believe future foundational OO PL research, to further
its relevance to mainstream OOP, should be based less on structural
models of OOP and more on nominal ones instead.

\bibliographystyle{plain}

\end{document}